\documentclass[10pt,conference]{IEEEtran}
\usepackage[utf8]{inputenc} 
\usepackage[T1]{fontenc}
\usepackage{url}

\usepackage{breakurl}
\usepackage{ifthen}
\usepackage{graphicx}
\usepackage[T1]{fontenc}

\usepackage{enumitem}
\usepackage[cmex10]{amsmath}
\usepackage{amssymb}
\usepackage{multicol}

\usepackage{cite}
\usepackage{amsthm}
\usepackage{textcomp}
\usepackage{siunitx}
\usepackage{color}
\usepackage{cases}
\usepackage[font={small}]{caption}
\usepackage{epstopdf}
\usepackage{graphicx}
\usepackage{caption}
\usepackage{verbatim} 
\usepackage{bbm}
\usepackage{mathtools}

\newtheorem{theorem}{Theorem}
\newtheorem{lemma}{Lemma}

\newtheorem{remark}{Remark}

\theoremstyle{definition}

\newtheorem{definition}{Definition}

\theoremstyle{remark}

\DeclareMathOperator{\argmin}{argmin} 
\newtheorem*{theorem*}{Theorem}

\usepackage[ruled,vlined]{algorithm2e}
\SetAlFnt{\footnotesize} 
\setlist[description]{style=multiline}

\begin{document}
\sloppy
\title{Carbon-Neutralized Task Scheduling for Green Computing Networks}
    
\author{Chien-Sheng Yang, Chien-Chun Huang-Fu and I-Kang Fu\\
MediaTek Inc. \vspace{-3mm}}

\maketitle

\begin{abstract}
Climate change due to increasing carbon emissions by human activities has been identified as one of the most critical threat to Earth. Carbon neutralization, as a key approach to reverse climate change, has triggered the development of new regulations to enforce the economic activities toward low carbon solutions. Computing networks that enable users to process computation-intensive tasks contribute huge amount of carbon emissions due to rising energy consumption. To analyze the achievable reduction of carbon emissions by a scheduling policy, we first propose a novel virtual queueing network model that captures communication and computing procedures in networks. To adapt to highly variable and unpredictable nature of renewable energy utilized by computing networks (i.e., carbon intensity of grid varies by time and location), we propose a novel carbon-intensity based scheduling policy that dynamically schedules computation tasks over clouds via the drift-plus-penalty methodology in Lyapunov optimization. Our numerical analysis using real-world data shows that the proposed policy achieves $54\%$ reduction on the cumulative carbon emissions for AI model training tasks compared to the queue-length based policy.

\end{abstract}

\section{Introduction}

Global warming caused by excessive emissions of carbon dioxide (e.g., burning fossil fuels for electricity generation) is the main driver to climate change, which has posed a significant threat to human society. To limit global warming, the most essential approach is via carbon neutralization, i.e., compensate carbon emissions by acquiring carbon offsets. Although the offsetting mechanisms for trading carbon credits (e.g., UN Carbon Offset Platform~\cite{UNoffset}) have been widely adopted globally, it has been shown that such mechanisms have limitations to effectively reduce the emissions~\cite{netzero}. To achieve carbon neutrality, it is important to reduce the carbon emissions in the first place rather than offset them later.


Due to recent advancements in computing networks that enable users to offload computation-intensive tasks to clouds, service demands for computing and communication resources in networks have been dramatically rising since 2010~\cite{Masanet2020RecalibratingGD}. Thus, carbon emissions due to increasing energy consumption in computing networks become a matter of concern. To reduce their carbon footprint and limit their environmental impacts, clouds have been pushed to use more renewable energy, e.g., Amazon AWS's goal of $100\%$ renewable energy by 2025~\cite{AWSrenewable}. 


Electricity generation is from energy sources (e.g., gas, coal, wind energy) with different levels of carbon emissions. In particular, due to the highly variable and unpredictable nature of renewable energy sources (e.g., solar energy), carbon intensity (i.e., average carbon emissions per unit of energy consumption) of electricity grid varies considerably by time and location~\cite{khan2019temporal,callaway2018location}. Thus, to guarantee the reduction of carbon emissions in computing networks, there is a critical need to design a task scheduling policy for networks, which accounts for temporal and spatial dimensions of energy sources.


In this paper, we consider the problem of task scheduling over computing networks with focus on the reduction of carbon emissions. More precisely, the considered computing network model is composed of an edge server and multiple clouds, in which the offloaded tasks arrive to the edge dynamically and then are dispatched to clouds accordingly. The edge server is responsible for sending data of tasks to clouds, and the energy consumption of edge server depends on which type of tasks it is sending. Each cloud is responsible for processing tasks, and the energy consumption of a cloud depends on which type of tasks it is processing. Subject to the energy consumption constraints, we assume that the edge server and each of clouds use different electricity grid, i.e., have different carbon intensity. To design an efficient scheduling policy that minimizes the carbon emissions from computing networks, we aim at exploiting the workload flexibility in both when and where the computation tasks are executed.


To analyze the carbon emissions from the network, we first propose a novel virtual queueing network model that captures the communication and computing procedures in the network. Then, in order to adapt to varied carbon intensity of electricity grids, we introduce the drift-plus-penalty methodology of Lyapunov optimization~\cite{neely2010stochastic}, whose idea is to minimize an upper bound on the drift-plus-penalty term (i.e., a linear combination of drifts and the carbon emissions with positive sign) at each time slot. Under the i.i.d assumption of the number of arriving tasks and the carbon intensity of edge and clouds, the introduced drift-plus-penalty methodology provides the guarantee on mean-rate stability of queues and achieves time-average carbon emissions arbitrarily close to optimal. 

The minimization for the upper bound of drift-plus-penalty in our scheduling problem, however, is shown to be a NP-hard unbounded Knapsack problem. Through the greedy approach for minimizing the upper bound of drift-plus-penalty, we propose an efficient dynamic carbon-intensity based scheduling policy. Using the randomly generated data and the real-world data (from National Grid ESO~\cite{nationalgrideso}) of carbon intensity, we conduct the numerical studies for the case of AI model training tasks. We show that the proposed carbon-intensity based policy can significantly outperform the queue-length based policy in terms of cumulative carbon emissions, while ensuring the mean-rate stability of queues. 

\vspace{3mm}
\noindent \textbf{Related Works:} We provide a literature review that covers the works of online scheduling and carbon-aware network.

The online scheduling problem aims to dynamically schedule jobs that arrive to the network according to a stochastic process. One of the main goals is to find a throughput-optimal policy~\cite{eryilmaz2005stable}, i.e., a policy that stabilizes the network, whenever it can be stabilized. For instance, Max-Weight type policy~\cite{tassiulas1992stability} has shown to be throughput-optimal for wireless networks, flexible queueing networks~\cite{neely2005dynamic} and dispersed computing networks~\cite{yang2019communication}. Furthermore, Lyapunov optimization is a technique that minimizes drift-plus-penalty to ensure the network stability and the maximization of stochastic utility~\cite{neely2010stochastic}. 

Carbon-aware network has been widely investigated in recent years to mitigate the global warming issue due to the escalating carbon emissions. One of key approaches for the reduction of carbon emissions is to do task scheduling by considering the temporal and spatial dimensions of energy sources~\cite{aldossary2021towards,9770383,6877266}. Based on the information of carbon intensity,~\cite{aldossary2021towards} formulated a static scheduling problem for the resources usage and the placement of virtual machines via mixed-integer linear programming, and proposed a multi-level approach to minimize the carbon emissions of data centers.~\cite{6877266} proposed a Lyapunov-based algorithm for clouds that minimizes electricity cost and poses a limit on the carbon emissions. By delaying temporally flexible compute workloads based on the forecast of next day's carbon intensity,~\cite{9770383} introduced a Carbon-Intelligent Compute Management to reduce carbon footprint of clouds. To distinguish from these carbon-aware approaches, with an objective to minimize carbon emissions, our proposed policy decides when and where to execute computation tasks dynamically. Furthermore, without any a-priori statistical knowledge and future predictions, the proposed policy is only based on the current state of computing network, i.e., number of arriving tasks, number of waiting tasks and real-time carbon intensity.

\emph{Notation}: We denote by $[N]$ the set of $\{1,2,\ldots,N\}$ for any positive integer $N$. We denote by $\mathbb{N}^{0}$ the set of non-negative integers, i.e., $\mathbb{N}^{0} = \{0,1,\ldots\}$. 
\section{System Model} \label{sec:sys}
We consider a computing network in which there is an edge server connecting to some clouds. Users offload their computation tasks to the edge server in an online manner, and then each computation task is executed by one of clouds. In particular, the electricity grids of network generate carbon emissions, due to the energy consumption for providing services.

In the network, there is one edge server and $N$ clouds. We consider $M$ types of computation tasks, which are possibly offloaded to the system by users. We consider the system in discrete time (i.e., $t=0,1,\ldots$). Let $a_m(t)$ be the number of type-$m$ tasks that arrive to the edge at time $t$. For each type $m \in [M]$, we denote by $p^{\text{e}}_m$ the energy consumption incurred by the edge server for sending a type-$m$ task to one of clouds. We denote by $p^{\text{c}}_{m,n}$ the energy consumption incurred by cloud $n$ for processing an offloaded type-$m$ task. At each time slot, we assume that the edge server has constant energy constraint $P^{\text{e}}$ for communicating data to clouds, and each cloud $n\in [N]$ has constant energy constraint $P^{\text{c}}_n$ for processing tasks.

 Carbon intensity, defined as the amount of carbon emissions per unit of energy consumption (e.g., gCO2 per kW$\cdot$h) is used to estimate the amount of carbon emissions incurred by the computing and communication procedures in the network. We assume that the edge server and each cloud utilize different energy sources including non-renewables (e.g., fossil) and renewables (e.g., wind), which have variation in carbon intensity. Specifically, we denote by $C^{\text{e}}\left(t\right)$ the carbon intensity of grid utilized by the edge server at time $t$, and denote by $C^{\text{c}}_n\left(t\right)$ the carbon intensity of grid utilized by cloud $n \in [N]$ at time $t$.


\subsection{Problem Statement}
In the task scheduling problem of computing network, a scheduling policy determines the followings: 1) when each task is sent to one of clouds, 2) the destination of each task, and 3) when each task is processed. Concretely, we define the following terms to characterize a scheduling policy. We denote by $d_{m,n}(t)$ the number of type-$m$ tasks that are sent to cloud $n$ at time $t$, and denote by $w_{m,n}(t)$ the number of type-$m$ tasks that are processed by cloud $n$ at time $t$. That is, at time $t$, a scheduling policy determines an action which is composed of $\left\{ d_{m,n}(t) \right\}_{\forall m \in [M], \ \forall n \in [N]}$ and $\left\{w_{m,n}(t) \right\}_{\forall m \in [M], \ \forall n \in [N]}$. Let $P^{\text{e}}_{\text{total}}(t)$ be the total energy consumption by the edge server and $P^{\text{c}}_{n,\text{total}}(t)$ be the total energy consumption by cloud $n$, which can be written as follows:
\begin{align}
    P^{\text{e}}_{\text{total}}(t) & = \sum^M_{m=1}\sum^N_{n=1}d_{m,n}(t)p^{\text{e}}_m;\\
    P^{\text{c}}_{n, \text{total}}(t) & =  \sum^M_{m=1}w_{m,n}(t)p^{\text{c}}_{m,n}, \ \forall n \in [N].
\end{align}

An action is \emph{feasible} if the constraints on energy consumption are satisfied\footnote{We assume that the scheduled tasks will be successfully communicated (processed) if the energy constraint of edge server (cloud) is satisfied.}, i.e., 
\begin{align}
    P^{\text{e}}_{\text{total}}(t) & \leq P^{\text{e}}; \\
    P^{\text{c}}_{n, \text{total}}(t) & \leq P^{\text{c}}_n, \ \forall n \in [N]. 
\end{align}

We denote by $C(t)$ the carbon emissions of the computing network at time $t$. Based on carbon intensity $C^{\text{e}}(t)$ and $C^{\text{c}}_n(t)$ at time $t$, $C(t)$ can be written as follows:
\begin{align}
    C(t) = C^{\text{e}}\left(t\right)\cdot P^{\text{e}}_{\text{total}}(t) + \sum^N_{n=1}C^{\text{c}}_n\left(t\right) \cdot P^{\text{c}}_{n, \text{total}}(t). \label{eq:carbon}
\end{align}
\begin{definition}[Time-Average Carbon Emissions]
    Given carbon emissions $C(t)$ at each time $t$, the time-average carbon emissions, denoted by $\bar{C}$, is defined as follows:
    \begin{align}
        \bar{C} = \limsup_{T \rightarrow \infty}\frac{1}{T}\sum_{t=0}^{T-1}\mathbb{E}\left[C\left(t\right)\right].
    \end{align}
\end{definition}
Based on the above system model, our main goal is to design a scheduling policy that chooses a feasible action on both when and where the computation tasks are executed at each time to minimize time-average carbon emissions $\bar{C}$.

\section{Virtual Queueing Network Model}\label{sec:queue}

To analyze the resulting carbon emissions using a scheduling policy, we model a virtual queueing network that encodes the state of the computing network. Then, we introduce an optimization problem that ensures the minimization of carbon emissions and the mean-rate stability of queues.

\begin{figure}[t]
    \centering
    \includegraphics[width = 0.8\columnwidth]{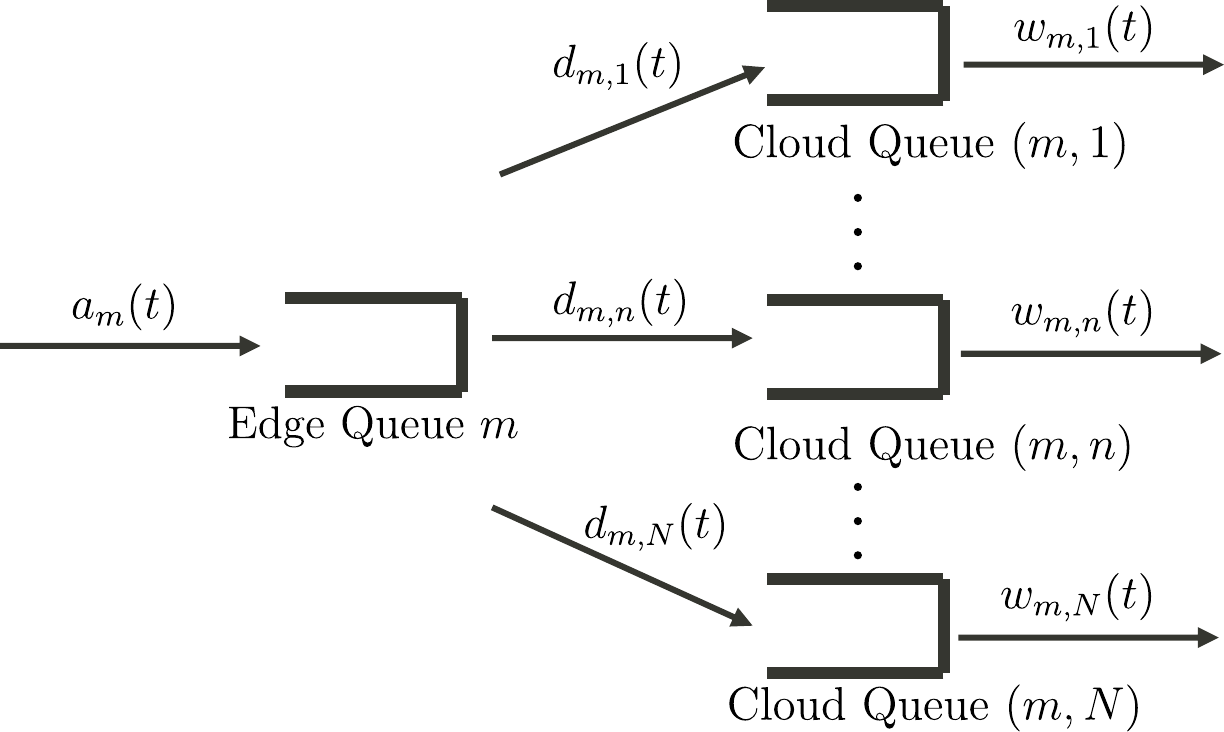}
    \caption{The illustration of proposed queueing network. For each task type $m \in [M]$, we maintain an edge queue $m$ and cloud queues $(m,n)$, $\forall n \in [N].$ At time $t$, $a_m(t)$ number of type-$m$ tasks arrive to edge queue $m$. Based on a scheduling policy, $d_{m,n}(t)$ number of type-$m$ tasks in edge queue $m$ arrive to cloud queue $(m,n)$, and $w_{m,n}(t)$ number of type-$m$ tasks depart from cloud queue $(m,n)$.}  
    \label{fig:queue}
\end{figure}

As shown in Fig.~\ref{fig:queue}, the proposed virtual queueing network consists of two kinds of queues, \emph{edge queue} and \emph{cloud queue}, which are modeled in the following manner:
\begin{itemize}[leftmargin=*]
    \item \textbf{Edge Queue}: We maintain one virtual queue called edge queue $m$ for type-$m$ tasks located in the edge server.
    \item \textbf{Cloud Queue}: We maintain one virtual queue called cloud queue $(m,n)$ for type-$m$ tasks processed by cloud $n$. 
\end{itemize}

We describe the dynamics of the virtual queues in the network. The type-$m$ tasks are sent to edge queue $m$ when arriving to the edge server. The tasks in edge queue $m$ are sent to cloud queue $(m,n)$ if the type-$m$ tasks are scheduled on cloud $n$ for processing. We denote by $Q^{\text{e}}_{m}(t)$ the length of edge queue $m$ and denote by $Q^{\text{c}}_{m,n}(t)$ the length of cloud queue $(m,n)$ at time $t$. We state the dynamics of the proposed queueing network as follows. For $\forall m \in [M]$, we have
 \begin{align}
    Q^{\text{e}}_{m}(t+1) & = \max (Q^{\text{e}}_{m}(t) - \sum^N_{n=1}d_{m,n}(t),0)+a_m(t). \  \label{eq:dynamics1}
    \end{align}
For $\forall m \in [M], \forall n \in [N]$, we have
    \begin{align}
    Q^{\text{c}}_{m,n}(t+1) & = \max(Q^{\text{c}}_{m,n}(t)-w_{m,n}(t),0)+d_{m,n}(t). \label{eq:dynamics2}
 \end{align}

Now, we introduce an optimization problem called \emph{carbon-aware queueing network planning problem (CQNPP)} that minimizes time-average carbon emissions $\bar{C}$ and stabilize all the queues in the virtual queueing network:

\noindent {\bf Carbon-Aware Queueing Network Planning Problem}
 \begin{align}
     & \ \min \ \ \bar{C}\label{eq:avg_carbon} \\
     s.t. & \ \lim_{T \rightarrow \infty}\frac{\mathbb{E}[Q^{\text{e}}_{m}(T)]}{T} = 0, \ \forall m \in [M]; \label{eq:queue_edge} \\
          & \ \lim_{T \rightarrow \infty}\frac{\mathbb{E}[Q^{\text{c}}_{m,n}(T)]}{T} = 0, \ \forall m \in [M], \ \forall n \in [N]; \label{eq:queue_cloud} \\
          & \ P^{\text{e}}_{\text{total}}(t)  \leq P^{\text{e}}; \label{eq:energy_edge} \\
          & \ P^{\text{c}}_{n, \text{total}}(t)  \leq P^{\text{c}}_n, \ \forall n \in [N];  \label{eq:energy_cloud} \\
          & \ d_{m,n}(t), w_{m,n}(t) \in \mathbb{N}^{0}, \ \forall m \in [M], \ \forall n \in [N] \label{eq:action}.
 \end{align}
 In CQNPP, \eqref{eq:queue_edge} and \eqref{eq:queue_cloud} indicate that we make each queue mean-rate stable; \eqref{eq:energy_edge}, \eqref{eq:energy_cloud} and \eqref{eq:action} define the space of feasible actions. The proposed CQNPP is a sequential decision-making problem, which is in general challenging to solve. 

\section{Carbon-Intensity Based Scheduling Policy}
 In this section, we introduce the drift-plus-penalty methodology in Lyapunov optimization~\cite{neely2010stochastic} to effectively minimize the carbon emissions and make queues mean-rate stable. 
Then, we design an efficient carbon-intensity based scheduling policy which dynamically decides "where" and "when" tasks are processed based on the current state of network, without any a-priori statistical knowledge and future predictions.
\subsection{Drift-Plus-Penalty Methodology}
We now introduce the drift-plus-penalty methodology for the proposed CQNPP. As a measure of congestion in virtual queues, Lyapunov function $L(t)$ is defined as follows:
\begin{align}
    L(t) = \frac{1}{2} \left(\sum^M_{m=1} Q^{\text{e}}_m(t)^2+ \sum^M_{m=1}\sum^N_{n=1}Q^{\text{c}}_{m,n}(t)^2\right). 
\end{align}
Then, we define the drift of Lyapunov function $L(t)$ as follows
\begin{align}
    \Delta(t) = L(t+1) - L(t).
\end{align}


To stabilize all the queues and minimize the carbon emissions, the key idea is to minimize the drift-plus-penalty, which is a weighted sum of drift and scaled penalty. Consider a non-negative number $V$, we formally define the drift-plus-penalty 
as $\Delta(t)+VC(t)$, where the penalty term at time $t$ is carbon emissions $C(t)$. Rather than directly minimize $\Delta(t)+VC(t)$ every slot $t$, we minimize an upper bound on this drift-plus-penalty expression. The following lemma provides an upper bound on the drift-plus-penalty.
\begin{lemma}[Drift Bound]\label{lemma:drift_bound}
    Suppose $a_m(t)$ is upper-bounded for all $m$ and all $t$. For any scheduling policy, drift-plus-penalty $\Delta(t)+VC(t)$ can be upper-bounded as follows
    \begin{align}
        \Delta(t)& +VC(t) \leq B+ \sum^M_{m=1}Q^{\text{e}}_m(t)a_m(t)\nonumber\\
        & +  \sum^M_{m=1}\sum^N_{n=1}\left(VC^{\text{e}}(t)p^{\text{e}}_m + Q^{\text{c}}_{m,n}(t)-Q^{\text{e}}_{m}(t)\right)d_{m,n}(t) \nonumber\\
              &+ \sum^M_{m=1}\sum^N_{n=1}\left(VC^{\text{c}}_n(t)p^{\text{c}}_{m,n}-Q^{\text{c}}_{m,n}(t)\right)w_{m,n}(t) \label{eq:upper_bound_final}
    \end{align}
    where $B$ is a constant such that 
\begin{align}
    &\sum^M_{m=1}a_m(t)^2 + \sum^M_{m=1}\left(\sum^N_{n=1}d_{m,n}(t)\right)^2 \nonumber \\
    &+\sum^M_{m=1}\sum^N_{n=1}d_{m,n}(t)^2 + \sum^M_{m=1}\sum^N_{n=1}w_{m,n}(t)^2 \leq 2B, \ \forall t. \label{eq:B}
\end{align}
\end{lemma}
The proof of Lemma~\ref{lemma:drift_bound} is provided in Appendix~\ref{proof_appendix_lemma}.
\begin{remark}
    We note that constant $B$ defined in ~\eqref{eq:B} must exist since $a_m(t)$ is assumed to be upper-bounded and $d_{m,n}(t)$ and $w_{m,n}(t)$ are subject to the constraints defined in \eqref{eq:energy_edge} and \eqref{eq:energy_cloud}. 
\end{remark}

At each time $t$, given number of arriving tasks $a_m(t)$, virtual queue-lengths $Q^{\text{e}}_m(t)$, $Q^{\text{c}}_{m,n}(t)$
 and carbon intensity $C^{\text{e}}(t)$, $C^{\text{c}}_n(t)$, a policy denoted by $\eta$ aims at choosing a feasible action that minimizes the upper bound defined in \eqref{eq:upper_bound_final}. This is equivalent to minimize
 \begin{align}
    & \sum^M_{m=1}\sum^N_{n=1}\left(VC^{\text{e}}(t)p^{\text{e}}_m + Q^{\text{c}}_{m,n}(t)-Q^{\text{e}}_{m}(t)\right)d_{m,n}(t) \nonumber\\
    + & \sum^M_{m=1}\sum^N_{n=1}\left(VC^{\text{c}}_n(t)p^{\text{c}}_{m,n}-Q^{\text{c}}_{m,n}(t)\right)w_{m,n}(t) \label{eq:upper_bound_equi}
 \end{align}
where $d_{m,n}(t)$ and $w_{m,n}(t)$ are subject to \eqref{eq:energy_edge}, \eqref{eq:energy_cloud} and \eqref{eq:action}. 

The following theorem shows that the theoretical guarantees provided by policy $\eta$ when the number of arriving tasks and the carbon intensity are i.i.d over time slots.
\begin{theorem}\label{thm}
    Suppose $a_m(t)$ is upper-bounded for all $m$ and all $t$. If $a_m(t), \forall m$, $C^{\text{e}}(t)$ and $C^{\text{c}}_n(t), \forall n$ are i.i.d over time slots, scheduling policy $\eta$ with a non-negative number $V$ that minimizes~\eqref{eq:upper_bound_equi} provides the following guarantees:
    \begin{itemize}
      \item \textbf{Performance Guarantee.} The achieved time-average carbon missions $\bar{C}^{\eta}$ satisfies 
       \begin{align}
            \bar{C}^{\eta} \leq \bar{C}^{\text{opt}} + \frac{B}{V}
        \end{align} 
        where $B$ is the constant such that \eqref{eq:B} is satisfied for all $t$, and $\bar{C}^{\text{opt}}$ is the infimum time-average carbon emissions achievable by any policy.  
        \item \textbf{Stability Guarantee.} All queues are mean-rate stable.
    \end{itemize}    
\end{theorem}
The proof of Theorem \ref{thm} follows the similar arguments in \cite{neely2010stochastic}, and we thus omit it due to the page limit.
\begin{remark}
 Theorem~\ref{thm} shows that policy $\eta$ achieves the time-average carbon emissions which deviates from the optimal value by no more than $\frac{B}{V}$.   
\end{remark}

Now, we show that the minimization of \eqref{eq:upper_bound_equi} can not be solved efficiently. Since the edge server and the clouds have independent constraints \eqref{eq:energy_edge} and \eqref{eq:energy_cloud}, minimizing~\eqref{eq:upper_bound_equi} can be decoupled into some independent optimization problems. For the edge server, we have the problem defined as 
 \begin{align}
     \min & \sum^M_{m=1}\sum^N_{n=1}b_{m,n}(t)d_{m,n}(t); \label{eq:op_1}\\
     s.t. & \ \sum^M_{m=1}\sum^N_{n=1}d_{m,n}(t)p^{\text{e}}_m \leq P^{\text{e}}; \label{eq:op_2}\\
     & \ d_{m,n}(t) \in \mathbb{N}^{0}, \ \forall m \in [M], \ \forall n \in [N]; \label{eq:op_3}
 \end{align}
 and for each cloud $n$, we have the problem defined as 
 \begin{align}
    \min & \sum^M_{m=1}c_{m,n}(t)w_{m,n}(t); \label{eq:op_4}\\
    s.t. & \ \sum^M_{m=1}w_{m,n}(t)p^{\text{c}}_{m,n} \leq P^{\text{c}}_n; \label{eq:op_5}\\
    & \ w_{m,n}(t) \in \mathbb{N}^{0}, \ \forall m \in [M], \ \forall n \in [N] \label{eq:op_6}
\end{align}
 where $b_{m,n}(t) = VC^{\text{e}}(t)p^{\text{e}}_m + Q^{\text{c}}_{m,n}(t)-Q^{\text{e}}_{m}(t)$ and $c_{m,n}(t) = VC^{\text{c}}_n(t)p^{\text{c}}_{m,n}-Q^{\text{c}}_{m,n}(t)$ are fixed numbers after knowing all the queue-lengths and carbon intensity at time $t$. 
 
 Since the problem defined in \eqref{eq:op_1} to \eqref{eq:op_3} aims at minimizing an objective function, the optimal solution requires that $d_{m,n}(t) = 0$ if $b_{m,n}(t) >0$. After dropping $b_{m,n}(t)d_{m,n}(t)$'s with $b_{m,n}(t)>0$ by setting $d_{m,n}(t) = 0$, the problem in \eqref{eq:op_1} to \eqref{eq:op_3} with remaining variables is an unbounded Knapsack problem which has been shown NP-hard~\cite{garey1979computers}. The similar arguments also hold for the problem defined in \eqref{eq:op_4} to \eqref{eq:op_6}.

\subsection{Description of the Proposed Policy}
 We propose a carbon-intensity based scheduling policy (see Algorithm~\ref{alg:carbon_policy}), whose idea is to greedily schedule tasks starting from the tasks with the most negative values contributed to \eqref{eq:upper_bound_equi} per energy unit. The proposed policy at each time $t$ is dominated by the sorting procedures, which can be done efficiently (with the complexity almost linear in $MN$). Now, we provide more details of the proposed policy:
 \begin{itemize}[leftmargin=*]
     \item \textbf{Edge Server:} For each $m$, we find $n_1(m)$ such that $VC^{\text{e}}(t)p^{\text{e}}_m + Q^{\text{c}}_{m,n_1(m)}(t)-Q^{\text{e}}_{m}(t)$ is the smallest among all $n$ (equivalent to find $n_1(m)$ such that $Q^{\text{c}}_{m,n_1(m)}(t)$ is the smallest among all $n$). Then, we sort the task types in increasing order of ratio $\frac{VC^{\text{e}}(t)p^{\text{e}}_m + Q^{\text{c}}_{m,n_1(m)}(t)-Q^{\text{e}}_{m}(t)}{p^{\text{e}}_{m}}$ (equivalent to sort the task types in increasing order of ratio $\frac{Q^{\text{c}}_{m,n_1(m)}(t)-Q^{\text{e}}_{m}(t)}{p^{\text{e}}_{m}}$). Subject to energy constraint $P^{\text{e}}$, the edge server sends as many as possible of type-$m$ tasks to cloud $n_1(m)$ with the smallest value of $\frac{Q^{\text{c}}_{m,n_1(m)}(t)-Q^{\text{e}}_{m}(t)}{p^{\text{e}}_{m}}$ while $VC^{\text{e}}(t)p^{\text{e}}_m + Q^{\text{c}}_{m,n_1(m)}(t)-Q^{\text{e}}_{m}(t)$ is negative.  
     \item \textbf{Cloud:} For each cloud $n$, we sort the task types in increasing order of ratio $\frac{VC^{\text{c}}_n(t)p^{\text{c}}_{m,n}-Q^{\text{c}}_{m,n}(t)}{p^{\text{c}}_{m,n}}$ (equivalent to sort the task types in decreasing order of ratio $\frac{Q^{\text{c}}_{m,n}(t)}{p^{\text{c}}_{m,n}}$). Subject to energy constraint $P^{\text{c}}_n$, the cloud $n$ processes as many as possible of type-$m$ tasks with the largest value of $\frac{Q^{\text{c}}_{m,n}(t)}{p^{\text{c}}_{m,n}}$ while the value of $VC^{\text{c}}_n(t)p^{\text{c}}_{m,n}-Q^{\text{c}}_{m,n}(t)$ is negative. 
 \end{itemize}
 

\begin{algorithm}[!t]    
    \SetAlgoLined
    \textbf{Input:} $V$, $M$, $N$, $P^{\text{e}}$, $P^{\text{c}}_n$, $p^{\text{e}}_m$, $p^{\text{c}}_{m,n}$\;
    \textbf{Initialization:} $d_{m,n}(t) = 0$, $w_{m,n}(t) = 0$\;
    \For{$t \leftarrow 0,1,\ldots$}{
    Observe $C^{\text{e}}(t)$, $C^{\text{c}}_n(t)$ and $a_m(t)$\;
    $n_1(m) \leftarrow \argmin_{n \in [N]}Q^{\text{c}}_{m,n}(t)$\;
    Sort: $\frac{Q^{\text{c}}_{1,n_1(1)}(t)-Q^{\text{e}}_{1}(t)}{p^{\text{e}}_1} \leq \cdots \leq \frac{Q^{\text{c}}_{M,n_1(M)}(t)-Q^{\text{e}}_{M}(t)}{p^{\text{e}}_M}$\; 
    $P \leftarrow P^{\text{e}}$\; 
    \For{$m \leftarrow 1$ \KwTo $M$}{
        \If{$\lfloor \frac{P}{p^{\text{e}}_m}\rfloor > 0$}{
            \eIf{$VC^{\text{e}}(t)p^{\text{e}}_{m}+Q^{\text{c}}_{m,n_1(m)}(t)-Q^{\text{e}}_{m}(t) <0$}{
                $d_{m,n_1(m)}(t) \leftarrow \min(Q^{\text{e}}_{m}(t), \lfloor \frac{P}{p^{\text{e}}_{m}}\rfloor)$\;
                $P \leftarrow P - \lfloor \frac{P}{p^{\text{e}}_m} \rfloor p^{\text{e}}_{m}$\;
    }{break\;}
        }
    }
   
    \For{$n \leftarrow 1$ \KwTo $N$}{
        $P \leftarrow P^{\text{c}}_n$\;
    Sort: $\frac{Q^{\text{c}}_{1,n}(t)}{p^{\text{c}}_{1,n}}\geq \cdots \geq \frac{Q^{\text{c}}_{M,n}(t)}{p^{\text{c}}_{M,n}}$\;  
    \For{$m \leftarrow 1$ \KwTo $M$}{
        \If{$\lfloor \frac{P}{p^{\text{c}}_{m,n}}\rfloor > 0$}{
    \eIf{$VC^{\text{c}}_n(t)p^{\text{c}}_{m,n}-Q^{\text{c}}_{m,n}(t)<0$}{
    $w_{m,n}(t) \leftarrow \min (Q^{\text{c}}_{m,n}(t), \lfloor \frac{P}{p^{\text{c}}_{m,n}}\rfloor$)\;
    $P \leftarrow P - w_{m,n}(t)p^{\text{c}}_{m,n} $\;
    }{break\;}
    }
    }  
    }
    Update $Q^{\text{e}}_{m}(t+1)$ and $Q^{\text{c}}_{m,n}(t+1)$ according to \eqref{eq:dynamics1} and \eqref{eq:dynamics2}
    
    }
     \caption{Carbon-Intensity Based Policy}\label{alg:carbon_policy}
    \end{algorithm}



\section{Numerical Analysis}
In this section, we demonstrate the impact of the proposed carbon-intensity based scheduling policy by simulation studies. We evaluate the effectiveness of the proposed policy in terms of the cumulative carbon emissions. We consider a network composed of an edge server and $5$ clouds. The edge server has energy constraint $P^{\text{e}} = 4000$ kW$\cdot$h, and each cloud $n$ has energy constraint $P^{\text{c}}_n = 30000$ kW$\cdot$h. We consider $M=5$ types of AI model training tasks on ImageNet~\cite{deng2009imagenet}, whose computation and communication consumption are summarized in Table~\ref{tb:1}.\footnote{The estimation of $P^{\text{c}}_n$ is based on the annual energy consumption of Google~\cite{google}. The estimation of $P^{\text{e}}$ is based on the assumptions: the bandwidth of $100$ GB/s for edge and the energy efficiency of $0.023$ kW$\cdot$h/GB for data transmission \cite{malmodin2016energy}. The estimation of $p^{\text{e}}_m$ and $p^{\text{c}}_{m,n}$ are based on the size of ImageNet dataset~\cite{deng2009imagenet} and the inference complexity of each model \cite{luo2020comparison} respectively.} For each $m$, $a_m(t)$ is randomly chosen from $\{0,1,\ldots,400\}$ at each time $t$.

 We compare the proposed policy with a queue-length based policy that makes decisions based on queue lengths: At each time $t$, the edge server sends as many as possible of the tasks that are located in the longest edge queues to the shortest cloud queues, and each cloud processes as many as possible of tasks located in its longest cloud queues. Then, we consider two scenarios for carbon intensity:
\begin{enumerate}[leftmargin =*]
    \item \textbf{Random:} At each time $t$, each of carbon intensity $C^{\text{e}}(t)$ and $C^{\text{c}}_n(t)$ is randomly chosen from $\{0,1,\dots,700\}$.  
    \item \textbf{Real World:} National Grid ESO~\cite{nationalgrideso} provides the regional carbon intensity data in the UK (per $30$ mins), where $6$ regions' data are used to represent the carbon intensity of the edge server and $5$ clouds.
\end{enumerate}
\begin{table}[t]
    \vspace{2mm}
    \centering
    \begin{tabular}{ |c|c|c|c| } 
    \hline
    Type & Model & $p^{\text{c}}_{m,n}$ (kW$\cdot$h)& $p^{\text{e}}_{m}$ (kW$\cdot$h)\\
     \hline
    $m=1$ & ResNet50 & 74 & 3.45 \\
    $m=2$ & InceptionV3 & 97 & 3.45 \\ 
    $m=3$ & DenseNet121 & 54 & 3.45 \\
    $m=4$ & SqueezeNet & 16 & 3.45 \\
    $m=5$ & MobileNetV2  & 5.8  & 3.45 \\ 
     \hline
    \end{tabular}
    \caption{Summary for the energy consumption of AI training tasks. It is assumed that clouds are homogeneous, i.e., $p^{\text{c}}_{m,1} = \cdots = p^{\text{c}}_{m,5}$.}
    \label{tb:1}
    \end{table}
\begin{figure}[t]
    \centering
    \includegraphics[width =0.85\columnwidth]{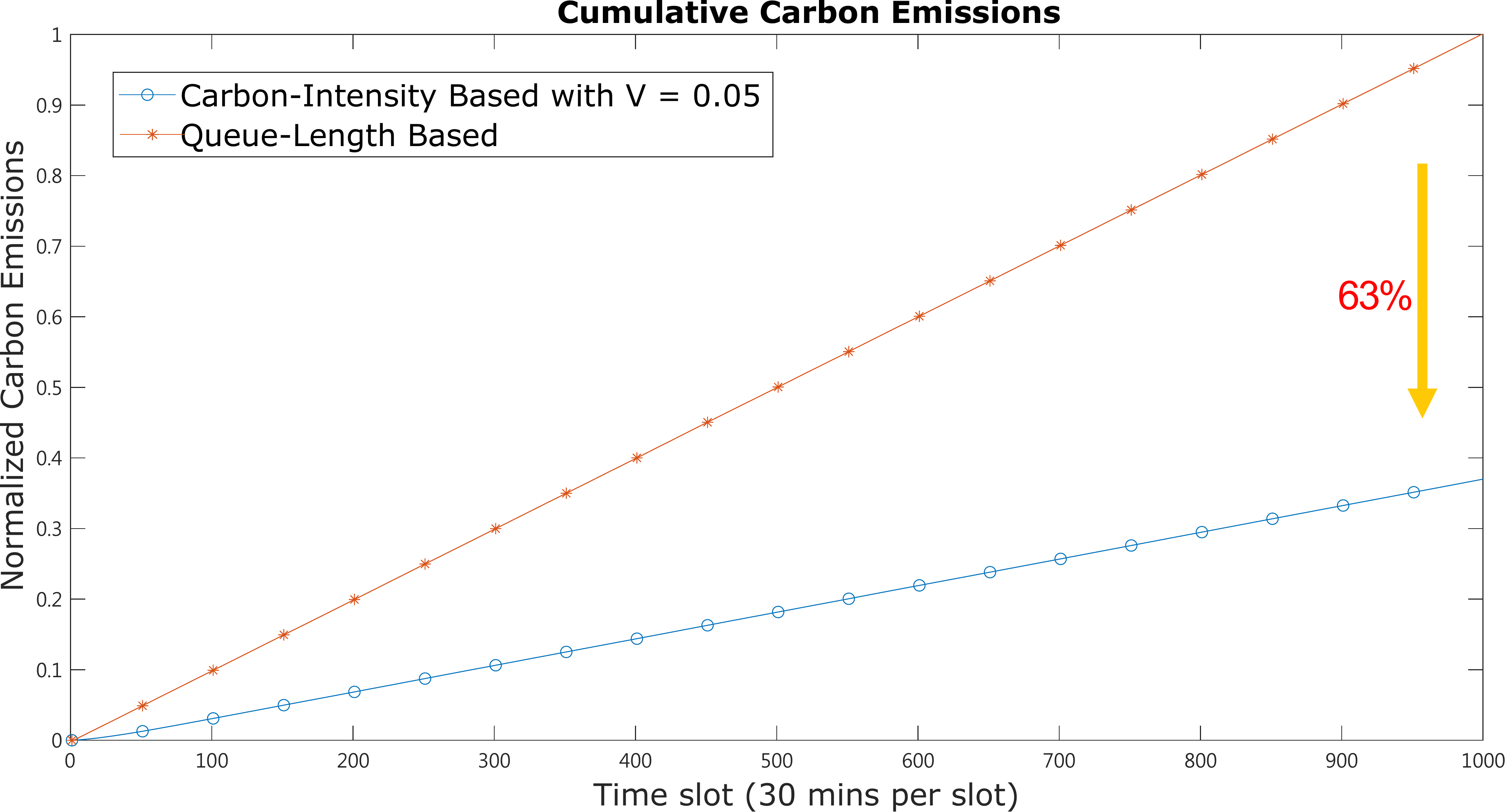}
    \caption{Numerical evaluations for cumulative carbon emissions (normalized) with the random carbon intensity.
    } 
    \label{fig:random_carbon}
\end{figure}
\begin{figure}[t]
    \vspace{2mm}
    \centering
    \includegraphics[width =0.85\columnwidth]{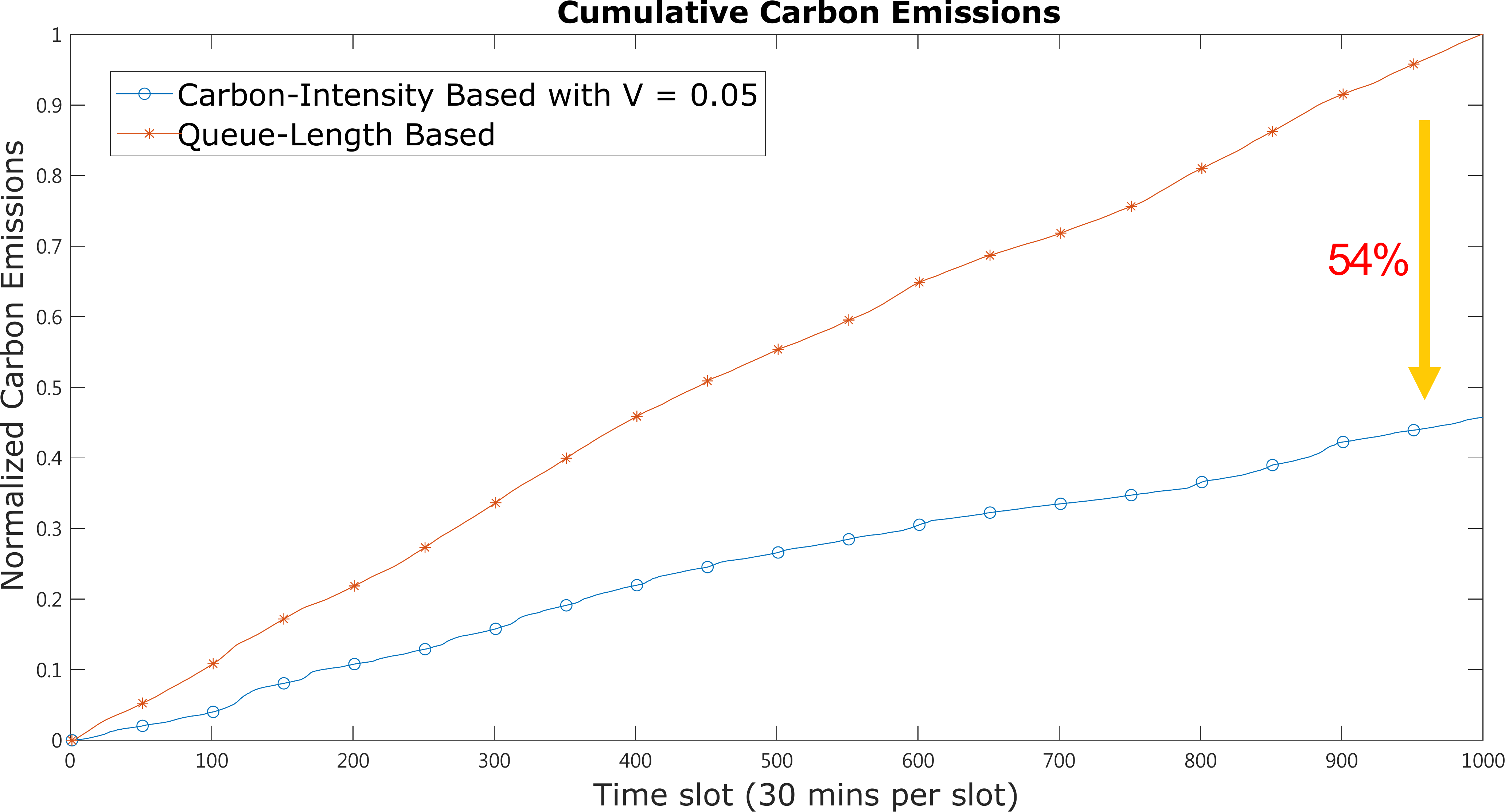}
    \caption{Numerical evaluations for cumulative carbon emissions (normalized) with the carbon intensity from National Grid ESO~\cite{nationalgrideso}.
    } 
    \label{fig:real_carbon}
\end{figure}
\begin{figure}[t]
    \centering
    \includegraphics[width =0.85\columnwidth]{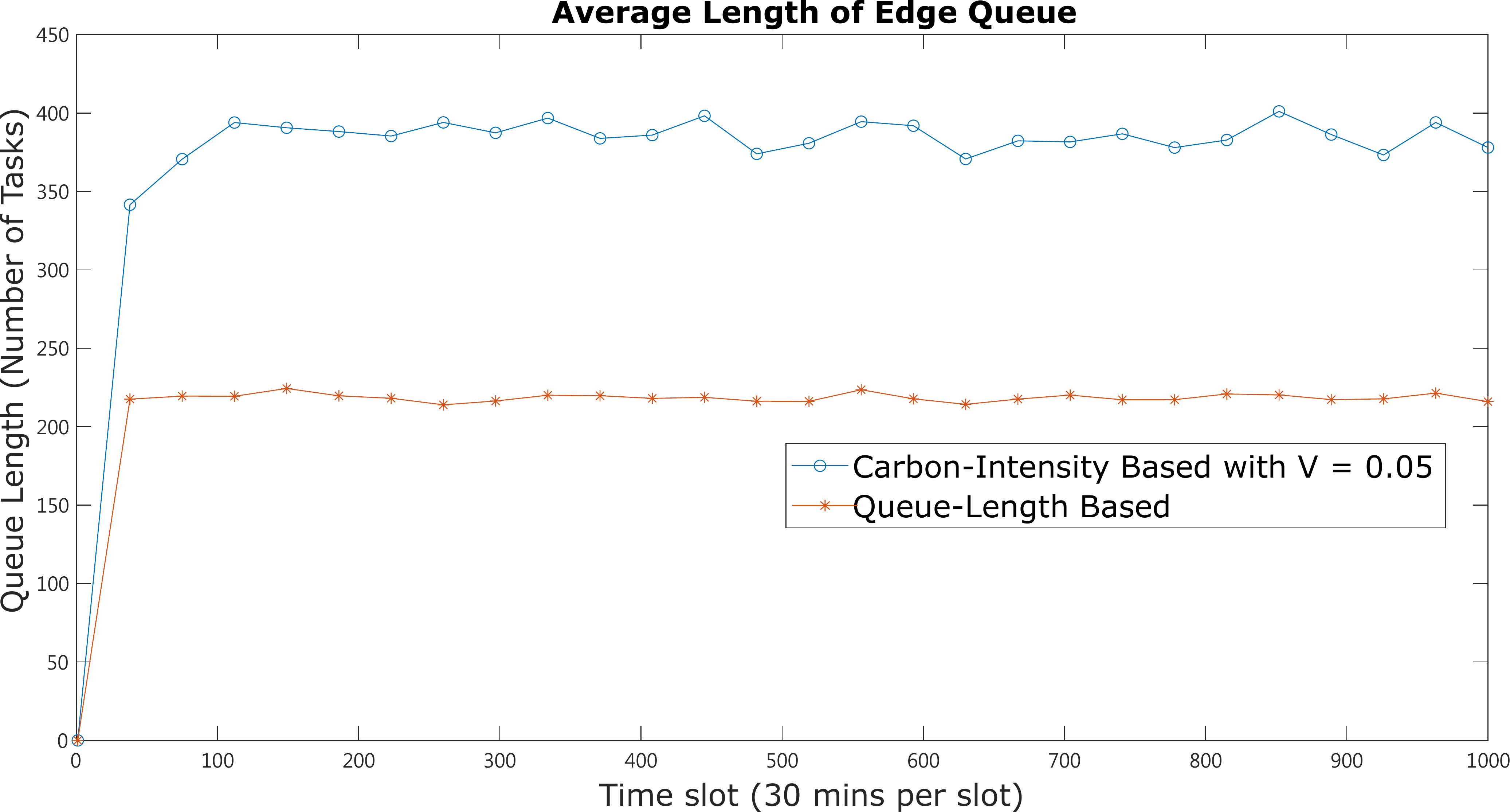}
    \caption{Numerical evaluations for average queue length with random carbon intensity.
    } 
    \label{fig:random_queue}
\end{figure}

Fig.~\ref{fig:random_carbon} and Fig.~\ref{fig:real_carbon} provide the normalized cumulative carbon-emissions comparison with the queue-length based policy.\footnote{The amount of carbon emissions will scale up in the sizes of energy consumption and energy constraint. Thus, we only focus on the normalized cumulative carbon emissions for the analysis.} Fig.~\ref{fig:random_queue} provides the comparison of average length of edge queue $m=1$ under the scenario of random carbon intensity. Then, we conclude the followings:
\begin{itemize}[leftmargin = *]
    \item For the random case, the proposed policy with $V = 0.05$ reduces the cumulative carbon emissions by $63\%$, and also ensures the mean-rate stability of queues. 
    \item For the real-world carbon intensity data, the proposed policy with $V = 0.05$ reduces the cumulative carbon emissions by $54\%$, which demonstrates the effectiveness of the carbon-intensity based policy in the real-world scenarios.\footnote{As indicated in \cite{siddik2021environmental}, the total carbon emissions attributed to data centers in
    2018 was $3.15 \times 10^7$ tons in the US. Thus, it is potential to reduce million tons of carbon emissions via our policy.}  
    \item Fig. \ref{fig:random_carbon} and Fig. \ref{fig:random_queue} indicate a tradeoff between carbon emissions and queueing delay provided by the underlying dirft-plus-penalty methodology. 
\end{itemize}

\section{Conclusion}


In this paper, we proposed a online carbon-intensity based scheduling policy for computing networks, which utilizes the temporal and spatial information of carbon intensity to effectively reduce carbon footprint of computing and communication procedures in the networks. Moreover, the leveraged drift-plus-penalty methodology provides the tradeoff between the reduction of carbon emissions and queueing delay. The numerical analysis in our paper demonstrates that the proposed scheduling policy can effectively reduce the overall carbon emissions by $54\%$ for AI model training tasks in the scenario of real-world carbon intensity. It is critical to take the carbon-related information into account when designing the communication and computation procedures of next-generation network in order to achieve the objective of carbon neutrality.

\bibliographystyle{ieeetr}
\bibliography{references}
\appendices
\section{Proof of Lemma~\ref{lemma:drift_bound}}\label{proof_appendix_lemma}
We first derive an upper bound on the sum of queue-length squares as follows:
\begin{align}
    & \sum^M_{m=1} Q^{\text{e}}_m(t+1)^2+ \sum^M_{m=1}\sum^N_{n=1}Q^{\text{c}}_{m,n}(t+1)^2 \label{eq:queue_square}\\
    & \leq  \sum^M_{m=1} Q^{\text{e}}_m(t)^2+ \sum^M_{m=1}\sum^N_{n=1}Q^{\text{c}}_{m,n}(t)^2 + \sum^M_{m=1}a_m(t)^2 \nonumber\\ 
    & + \sum^M_{m=1}\left(\sum^N_{n=1}d_{m,n}(t)\right)^2+\sum^M_{m=1}\sum^N_{n=1}\left(d_{m,n}(t)^2+ w_{m,n}(t)^2\right) \nonumber\\
    & +   2\sum^M_{m=1}Q^{\text{e}}_m(t) \cdot \left(a_m(t) - \sum^N_{n=1}d_{m,n}(t)\right) \nonumber \\
    & + 2\sum^M_{m=1}\sum^N_{n=1}Q^{\text{c}}_{m,n}(t)\cdot\left(d_{m,n}(t)-w_{m,n}(t)\right)  \label{eq:upper_bound1} 
\end{align}
where \eqref{eq:upper_bound1} follows from \eqref{eq:dynamics1}, \eqref{eq:dynamics2} and the inequality $(\max(a-b,0)+c)^2 \leq a^2+b^2+c^2 + 2a(c-b)$ for $a,b,c \geq 0$.

By rearranging equation~\eqref{eq:queue_square} and \eqref{eq:upper_bound1}, drift $\Delta(t)$ can be bounded as follows: 
\begin{align}
     \Delta(t) & \leq B + \sum^M_{m=1}Q^{\text{e}}_m(t) \cdot \left(a_m(t)-\sum^N_{n=1}d_{m,n}(t)\right)  \nonumber\\ 
     & + \sum^M_{m=1}\sum^N_{n=1}Q^{\text{c}}_{m,n}(t)\cdot \left(d_{m,n}(t)-w_{m,n}(t)\right) \label{eq:proof1}
\end{align}
where $B$ is a constant number defined in \eqref{eq:B}. By adding $VC(t)$ on both sides of \eqref{eq:proof1} with some rearrangements, we finally conclude the proof.


\end{document}